# The Future of CISE Distributed Research Infrastructure

A Community White Paper
03/08/2018

J.Aikat (RENCI/UNC Chapel Hill)
I.Baldin[1] (RENCI/UNC Chapel Hill)
M. Berman (BBN/Raytheon)
J. Breen (Utah)
R.Brooks (Clemson)
P. Calyam (Missouri)
J.Chase (Duke)
W.Chase (Clemson)
R. Clark (Georgia Tech)
C.Elliott (BBN/Raytheon)
J.Griffioen[1] (Kentucky)
D. Huang (ASU)
J.Ibarra (FIU)
T. Lehman (Maryland)
I.Monga (ESnet)
A.Matta (Boston University)
C. Papadopoulos (Colorado State)
M.Reiter (UNC Chapel Hill)
D.Raychaudhuri (Rutgers)
G. Ricart (US Ignite)
R. Ricci (Utah)
P. Ruth (RENCI/UNC Chapel Hill)
I.Seskar (Rutgers)
J.Sobieski (NORDUnet/GEANT)
K. Van der Merwe (Utah)
K.-C.Wang[1] (Clemson)
T. Wolf (UMass)
M. Zink (UMass)

---

[1] Correspondence authors ibaldin@renci.org, griff@netlab.uky.edu, kwang@clemson.edu

**Abstract:** Shared research infrastructure that is globally distributed and widely accessible has been a hallmark of the networking community. This paper presents an initial snapshot of a vision for a possible future of mid-scale distributed research infrastructure aimed at enabling new types of research and discoveries. The paper is written from the perspective of "lessons learned" in constructing and operating the Global Environment for Network Innovations (GENI) infrastructure and attempts to project future concepts and solutions based on these lessons. The goal of this paper is to engage the community to contribute new ideas and to inform funding agencies about future research directions to realize this vision.

## 1. Overview

An experimental testbed that can enable the research community to develop new network protocols, systems, and applications has long been an important goal. Around 2006, the U.S. National Science Foundation (NSF) sought to develop a Global Environment for Network Innovation (GENI). The current GENI testbed is an active research platform [2] that is used widely in the community. Entire research programs, such as the NSF Future Internet Architecture (FIA) [3], have relied on GENI or used the key concepts of such infrastructure [1, 4–7]. Similar testbeds have been developed and deployed in Europe and Asia.

Reflecting on the ten year journey with NSF GENI and observations of the myriad of activities and opportunities that stemmed from it, this white paper is an attempt to convey our opinions of the focus and strategies core to a successful future CISE distributed network research infrastructure that will continue to support and stimulate major discoveries while impacting CISE and other research communities.

## 2. Going Beyond GENI

Over the past decade GENI successfully proved itself an effective platform for research and instruction in a range of CISE focus areas such as:

- Distributed computing and edge clouds
- Networking (software defined networking, hardware, software, wireless, mobile, local, regional, wide area)
- Security
- Future clouds and smart cities

The successes resulted from several key accomplishments including:

- A federated approach for provisioning, accessing, and programming distributed cyberinfrastructure through standard abstractions, APIs, and security policies
- A suite of shared services, e.g., network stitching, experiment support, instrumentation
- Community-contributed resources -- e.g., network connectivity, VLANs, bandwidth, compute racks
- Community outreach, education and instructional materials, e.g., labs, tutorials, videos, workshops

What was constructed in the end was a modestly programmable network connecting deeply programmable distributed compute racks, supporting concurrent users performing isolated (sliced) experiments.

In spite of the successes, it also became clear that the GENI approach has not reached its full potential as many envisioned earlier during its ideation phase. Engineering decisions made under time-to-operation and manageability constraints led the community to a focus on a narrower set of resources and capabilities. A list of unmet desires includes:

- Insertion of researchers' own devices, instruments, research prototypes, and commercial cloud *from the edge* - teams able to purchase or manufacture their own equipment had no standard way of including this equipment into the testbed and interconnecting it using available testbed resources.
- Access to real world traffic - complete slice dataplane isolation from the commodity networks makes for more repeatable experiments, but also limited the scope and scale of the experiments.
- Programming the core network, with rich compute and storage - GENI largely relied on bandwidth-on-demand solutions from Internet2 and ESnet, with limited engagement with OpenFlow-enabled overlays, that required substantial effort to be used. Modern SDN solutions, including P4 would substantially enrich experimenter capabilities in this regard.
- Predictable network quality of service - while generally pervasive QoS must be traded against scale, the recent improvement in technologies used by campus operators and NRENs would make it possible to improve the situation for future experimenters.
- Testing innovative applications with real users - the infrastructure should allow hosting distributed applications open to real-world users
- Multi-domain networking experiments - current multi-domain capabilities are largely limited to creating isolated overlay slices using software virtualization. Instead infrastructure should make it possible to experiment with different architectural solutions in this space, while using real-world traffic and applications

Overall, GENI has demonstrated the power of a shared mid-scale infrastructure in enabling both technology-specific research (focusing on deep programmability of network and/or compute) and broader distributed applications research (focusing on distributed customized network and/or compute). Nonetheless, these unfulfilled needs reflect the opportunities to significantly boost research in a larger CISE community with a future CISE shared distributed mid-scale infrastructure. In this paper we present a vision of a future research infrastructure that incorporates past lessons from GENI's successes and limitations, and at the same time explores new directions that have the potential to open up entirely new areas of research and support a broad area of research, expanding the research community beyond that supported by GENI.

When considering a future research infrastructure there are many goals to balance against one another. The main objective of this document is to envision the features of a new infrastructure that will enable and support new areas of research -- research areas that are not well supported by other existing infrastructures or mechanisms. It should be noted that industry is often able to offer commodity services, and a future infrastructure should not duplicate -- but rather should complement -- commercial services when they are available.  Indeed, many emerging research areas will benefit greatly from the ability to explore questions at the boundary between production and research services.   While considerations such as the cost, organizational/management structures, and the developmental steps needed to achieve such an infrastructure are important, they are not the focus of this document and would need to be explored in follow-on efforts.

## 3. Key Premises

In order to motivate our vision, this section presents some possible research areas that the proposed infrastructure  can help enable, lists the various stakeholder groups that must be brought together to realize this vision, and finally formulates some key requirements of the proposed infrastructure.

## 3.1. Enabled Research Areas

We believe the envisioned infrastructure will realize the same multiplicative effects that GENI has achieved for its constituencies across a broader range of CISE researcher communities and research problems.  While the research communities that can be served by this infrastructure are quite diverse, we identify example areas whose needs may not be served by other existing infrastructure and would be significantly advanced by making this infrastructure available to the research community. These include network architecture research focusing on hybrid clouds, edge clouds, multi-domain networks, and SDXs; CPS, edge clouds, IoT and some mobile application research, especially in combination with PAWR platforms - an area acquiring ever-growing importance due to the rapid proliferation of IoT and smart devices; autonomous network management based on machine learning that treats the network itself as a 'big-data' instrument - a new area that is rich with research opportunities, and, finally, security research focusing on distributed security architectures and enabling the deployment of these solutions in an environment with real traffic and real users. All these areas will lead to transformative research that will reshape the way we interact with cyber-infrastructure, making it more pervasive, secure and autonomous.

In addition this infrastructure will benefit many types of domain research and provide for cross-fertilization opportunities between CISE and other domain researchers. We can imagine the infrastructure fulfilling multiple purposes depending on the stage and scale of the research effort: from validating research hardware and software, to testing scaling properties based on real-world user traffic, to deploying production-grade long-lived services that result from the research.

**Hybrid and Multi-domain Network and Cloud Architectures:** It is important to investigate placement and performance of computation and storage and switching equipment within the network and associated network resource control and management techniques in single- and multi-domain environments. GENI approach to this has been to creating software-based instances using OpenFlow overlays, limited in scale and performance. Having dedicated in-network (as opposed to the edge -- as in GENI) computation, storage and switching controllable by users, and attached to both edge clouds as well as NSF Clouds to generate workloads should allow this research to flourish.

In addition, some of the solutions can be transitioned to practice and tested under a variety of scenarios by allowing the exchange of traffic between experimental long-lived slices and production campus and global networks, at times using researcher-provided equipment.

It should also be noted that this infrastructure can play a crucial role in further development of end-to-end resource virtualization concepts. While the industry is actively working on standardization (Open Network Automation Platform (ONAP) and the Open Source MANO (OSM)), the lifecycle management of VNFs in a wireless context is an open research problem that is generating a lot of interest. Evolution drivers like coordination, centralization and virtualization are viewed as central tenants of any future (5G and beyond) distributed cloud based radio access network ("Cloud RAN") architectures. Any future CISE infrastructure should be instrumented to support experimentation with the range of possible RAN functional splits and implementations.

Interfacing with the upcoming PAWR testbeds would allow the integration of a wireless edge into experiments, which should provide significant value to PAWR experimenters interested in the interface between wireless edge and clouds. Future mid-scale infrastructure should plan to integrate existing GENI, CloudLab and PAWR resources to create a general framework for network, cloud and edge cloud research.

**CPS, SmartGrid and Smart Cities**: The exploration of cyber physical systems will require new capabilities for interacting with IoT-related devices as well as collecting large amounts of data and real-time control information from these sensors and devices located in the physical world. This will allow experimenters to perform intelligent data analysis and feedback of sensor data and control information, which can have dramatic impact on the effectiveness of novel integrated applications.

The integration of edge and core-based storage capabilities within experimental infrastructure will allow experimenters to study the emergence of novel architectures in this space by fusing and processing large amounts of data at various points in the network and creating different types of control algorithms, while studying their scalability and security.

This type of research requires reaching to campus or municipal networks and having substantial resources available to build scaled up systems that can be run as persistent services to evaluate long-term performance and e.g. economics while helping transition them into practice.

**Autonomous Network Control and Measurements:** New control structures will turn the network into a measurement platform for data-driven research, creating new types of autonomous control techniques, transforming how network and cloud infrastructures are deployed and operated. There is a significant and rising interest in this field, especially in cross-layer network control. However existing experimental facilities lack the scale and measurement capabilities to allow researchers to implement full-scale systems that can be evaluated over long periods of time with real user traffic. The proposed infrastructure will have the resources, connectivity and shared measurement capabilities/data to support this new research.

**Security Research:** Exploration of novel security approaches will be enabled by providing a platform in which reconfigurable widely distributed computational resources are configured with secure links that attach to secure systems and platforms at the campus edge, with suitable authorization and traffic control. Campuses may add interfaces that allow researchers with suitable authorization to sample traffic or direct traffic for specific applications and platforms through their security appliances, as mandated by campus policy or desired by the researchers. The proposed infrastructure will extend deep into a campus network, reaching researcher and user desktops, opening the door to experimentation with new security technologies on hosts or with users in ways that existing infrastructures rarely support. The wide geographic distribution of available resources will permit shifting of virtualized security functions within the substrate, thus allowing new security architectures and solutions to emerge. The researchers will create required target environments by combining resources from various infrastructure components, including NSF Clouds for larger scale experiments.

An interesting point of convergence is IoT security research, that simultaneously requires access to realistic IoT deployments, created by researchers on campuses, a programmable network capable of intelligently steering IoT data and substantial cloud processing capabilities to operate on this data.

The above enabled research areas are only a subset of the potential areas that could be enabled by a future distributed CISE research infrastructure. Moreover, one can imagine a wide range of science applications and domains (e.g., High Performance Computing applications) leveraging the infrastructure in novel new ways to make break-through scientific discoveries.

Thus, we envision the CISE distributed research infrastructure to be able to meet the needs of leading-edge CISE research (and possibly other NSF research), and to continue to evolve with new capabilities as informed by the community. This calls for a multi-faceted approach that consists of:

- A capable and stable Core Infrastructure
- A flexible Edge Infrastructure reaching into campuses
- A method to federate or otherwise integrate and manage external equipment, facilities, and external users of research applications
- An operation and support framework anchored in campus IT and Research & Education Networks (REN).

While other platforms suitable for experimentation exist in industry and academia (e.g. NSF and commercial clouds), those serve to complement, rather than compete with the envisioned infrastructure. Their main advantage is massive scale concentrated in a small number of geographically distributed locations. Their disadvantages are lack of wide-area network programmability, lack of fine geographic diversity required for low-latency applications and difficulty reaching local resources and users, as the cloud model presumes aggregation of resources, thus leaving out a broad and important range of research areas. The envisioned infrastructure is intended to address those shortcomings by providing user programmability, interfacing with campus and national resources and providing the ability to bring in real users and traffic onto experimental infrastructure. Together with the larger cloud resources, campus and national cyber resources it will create a rich experimental ecosystem that will support various types of systems and domain research for years to come.

## 3.2. Stakeholders and requirements

The envisioned future infrastructure will serve as a focal point that requires bringing together different types of stakeholders in order to make it a success:

- *Resource Contributors:* The testbed will consist of a variety of resources with many being contributed by universities (researchers and CIOs), cities, companies, and network providers. It is important that these resource/infrastructure owners see value in contributing/attaching their resources to the infrastructure. Importantly resources can be contributed not just by campuses or NRENs, but by research teams or their industry partners (Bring Your Own Research Equipment or BYORE).
- *Experimenters:* The testbed should be usable by researchers experimenting with all layers of the protocol stack, ranging from the physical layer (which will likely be enabled by new devices at the edge, wireless and/or optical technologies), the network layer (experimenting with new network architectures), the transport layer (new protocols), and uppers layers (e.g., new application protocols/interfaces), and supports both control plane and data plane experiments. The testbed should also provide operational data at scale, and provide "real traffic" (even if it is recorded) with controllable levels of privacy.
- *Instructors:* Like GENI, we expect the new testbed to be heavily used by instructors in their classes. However, going beyond GENI, we expect to see a much wider range of classes and application domains using it for instruction and training, with fewer limitations to the size and number of classes utilizing the infrastructure.
- *Users:* While GENI largely served network researchers and educators, the new infrastructure should be designed to support novel new applications that -- while being experimental -- are deployed as consumer-grade production application with the goal of attracting real world users who will actually use the application in their regular line of work.
- *Network service providers:* Regional and national network providers, also providers of commercial network services: Their interest and participation are vital to creating and maintaining the infrastructure, as well as for transitioning the results of research efforts into practice. Their infrastructures complement experimental infrastructures and provide additional resources that can be linked to experiments.
- *Industry and federal government representatives:* interested either in engaging the platform for advancing their research goals alongside traditional experimenters, using the applications deployed on the platform or in transitioning the results of the research into practice.

### 3.3. Key Requirements

To serve the broad research agenda and to attract stakeholders from such a wide range of groups will require connectivity that reaches them. GENI did an excellent job connecting a nontrivial number of campuses (or more specifically campus GENI racks) together. While this was a major accomplishment in itself, the proposed infrastructure must have a far more pervasive connectivity footprint if it is to reach all the potential stakeholders and beneficiaries.

First, it must extend to a campus network and beyond - reaching researcher and user desktops. While this may require some engineering work, the key here is that connectivity just "works" with little or no effort from end users. In other words, mechanisms that reach deep into campus networks need to be designed as first class connections; not potential add-ons.

Second, the infrastructure must extend to new types of research instruments ranging from mobile devices, to sensor devices, to control devices, to a wide range of IOT devices. The researchers must be allowed to BYORE and the network connectivity should treat connections to such devices as first class connections that are an integral part of the overall network topology. Those instruments can then be connected into a variety of experiments using functionality built into the platform (e.g. SDX). Moreover, the infrastructure should be designed to handle large and diverse sets of data (streaming data, long-term data storage, etc.) and data analysis (e.g., deep learning hardware) needed to support these types of devices and instruments.

Third, the infrastructure must extend to existing Internet and cloud services. The network connectivity should extend directly up to these services. This includes connectivity to national/public research facilities, like NSF Cloud and XSEDE, commercial cloud infrastructures, like those offered by Amazon, Microsoft and Google, and emerging national network research platforms such as PAWR. It must also provide access to smaller clouds that are widely distributed and provide low-latency computation and storage resources for mobile applications.

Fourth, the infrastructure must extend to international testbeds to enable international research collaboration and experimentation, this includes ongoing and emerging EU and APAC efforts that go beyond GENI. Future CISE and comparable international distributed research infrastructure should be accessible to a globally distributed community of researchers to jointly work on architectural principles and mechanisms for supporting advanced efficient resource management and topology embedding in distributed networked infrastructure environments targeted at experimentation. Connections to developing countries should be pursued, as they are qualitatively different in architecture with a predominantly wireless last mile and distinct problems.

While supporting real users and researchers simultaneously requires careful engineering, sufficient resource provisioning, operational management, and operational security, there is a significant value in providing a comprehensive connectivity that reaches in all directions. This connectivity will enable the envisioned infrastructure to attract real world users from the many science communities. It will connect those users with cutting-edge applications and promote cross-fertilization among domain science disciplines and CISE research community.

The infrastructure must provide extensive introspective capabilities in the core and at the edge to help understand the behavior of the infrastructure itself (e.g., monitoring and measurement capabilities that allow researchers to directly observe the performance and power consumption of different elements). In addition, accurate time measurement will be critical for time and latency sensitive applications and services, as well as for real-time based resource scheduling. Some of these capabilities can be developed by building on experiences learned, and features developed in, GENI. Other capabilities will require novel approaches that result in support for more or extended resource types, novel measurements and better scalability.

## 4. High-level Architecture and Enabling Features

With our objectives in mind, in this section we outline the necessary structure and features of the proposed infrastructure without fully specifying the architecture. We envision the CISE infrastructure to consist of two major components with an emphasis of their ability to interconnect with each other and other research infrastructure and campus researchers: (1) *Future Core Network* (FCN) will permit experimentation with novel and capable in-networking processing and storage concepts, and (2) *Future Edge Cloud* (FEC) which will have the resources and ability to reach users and applications at the edge and provide the experimental workloads that traverse the FCN. The proposed FCN/FEC network will form a programmable substrate interconnecting a variety of national and international research infrastructures as well as public cloud providers who are playing an increasing role in supporting national research objectives. The FCN/FEC will leverage past

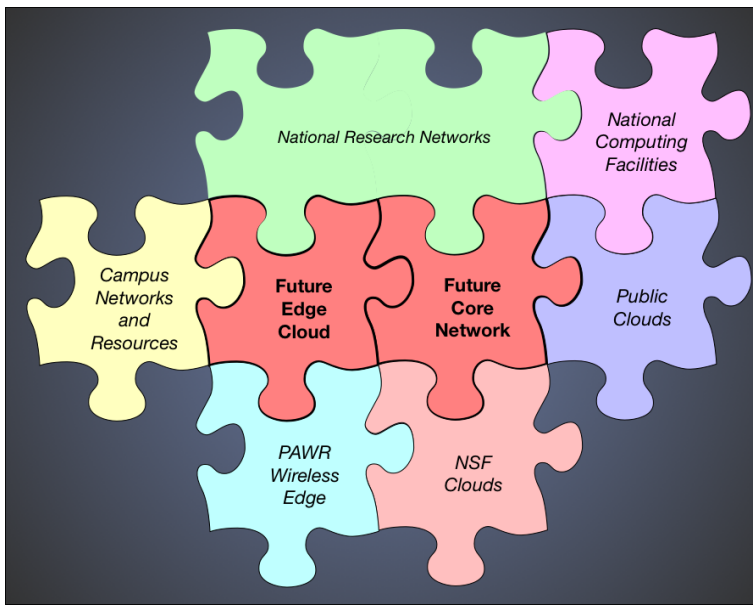

investments and key results from past projects including GENI, NSF Cloud, IRNC and CC* programs.

An orchestration system will provide access and ensure interoperability and consistency across the FCN/FEC infrastructure. It will provide appropriately authorized APIs for different levels of resource abstractions allowing applications to manage resource allocation on behalf of researchers and simplifying the interactions between experimenters and the system.

FCN will be built as an experimental core network centered around the concept of in-network compute and storage. In addition to both legacy and SDN-enabled switching and routing capabilities, it will have significant sliverable and programmable computing and storage capabilities that will allow the network to not only forward the data, but, critically provide the capacity for processing (e.g., indexing, filtering, converting, learning/reasoning) and buffering/storage, thus permitting experimentation with the concept of a hybrid infrastructure that has varying cloud and network properties. Modern SDN capabilities will provide significant degree of programmability of configuration and behavior and support experimentation with various topics in SDN as well as with in-network NFV, all while interconnecting elements of national cyber-infrastructure (production and experimental) to allow for a wide spectrum of experiments with technologies at different levels of scale and production readiness. FCN will provide mechanisms and policies for easy on-demand peering with commercial providers to allow experimenters exchange traffic with production infrastructure.

FEC will be modeled after the current GENI architecture in the sense of having widely distributed programmable resources that are deeply embedded into campus infrastructure of US universities and labs. By providing significant compute and storage capacity at campus edges it will permit dual experimental/production use with e.g. big-data in wide-area-network settings allowing to instantiate edge functions, like NFV, CDN, firewall/IDS and others. Critically, from the very start FEC sites will serve as bridges between FCN capabilities and other campus resources, that experimenters may want to include, such as Science DMZs and researcher-provided instruments.

Exchanging traffic will be accomplished by allowing controlled programmatic peerings between experimental dataplanes and commodity Internet implemented as SDXs. This will be accomplished by deploying peering capabilities at campus and core infrastructure level to guarantee diversity of resources and scale. Many campuses, as a result of CC* program investments now have high levels of programmability in their networks that will permit steering selected traffic into the testbed. To guarantee positive outcomes, the proposed infrastructure will require the hosting campuses to open their networks and provide the capability for bringing in local campus users. Similarly, by working with commercial and R&E network providers, future GENI will enable experimenters to peer their infrastructure with the core network infrastructure.

To support long-running experiments deploying 'academic-production' grade services, the proposed infrastructure (a) will have sufficient reliability of components, (b) will rely on the ability to steer production traffic in- and out-of the experimental slices and (c) will host an app-store of service-oriented appliances allowing for a quick deployment of functionality at large scale.

Finally, by supporting BYORE approach, experimenters will be able to launch persistent services using a mix of existing FCN/FEC equipment and the equipment they are able to acquire themselves, installed on their behalf within FCN/FEC infrastructure.

As mentioned in Section 3.2, the FCN will connect to public clouds, NSF Clouds as well as to national computing facilities, like XSEDE, DOE resources, leadership computing facilities to enable research in topics enabling future distributed computational domain science research, like e.g. construction of science super-facilities.

While FCN is intended to provide the experimental connectivity between elements, for experiments requiring reliable production-grade connectivity, there will be an option of using traditional bandwidth-on-demand Layer 2 services from national R&E network providers.

To simplify experimentation and make the experiments repeatable, both infrastructures will provide multi-layered monitoring capabilities and a well-established 'experiment profile' feature allowing to construct, preserve and share configurations of different experiments. Also, 'sandbox' areas will provide a way to debug aspects of experiments at a smaller scale, but with a greater level of visibility and control of the resources, prior to scaling them up to the entire testbed.

## 5. Conclusion

Past experience deploying the GENI testbed has provided the research community with a wide range of valuable insights into both the design of a testbed and its usage. While GENI was a significant step forward and much has been learned from GENI, many research needs went unmet in GENI. The key to a future CISE distributed research infrastructure is to understand those unmet needs and ensure they are considered in any future designs. The goal of this white paper has been to initiate the discussion regarding the key features that should be supported in a future testbed. It is expected that this is only the first step in such a conversation. As the discussion continues, additional constraints such as funding limitation and cost/benefit analysis will need to enter in which will likely result in realistic decisions being made about the features and capabilities that ultimately become goals of the new infrastructure. Some of the most challenging hurdles may ultimately be organizational in nature -- requiring buy-in from all the stakeholders and domains that will need to join and contribute to the new infrastructure. While these challenges are quite real, it is our belief that our experience with GENI has ideally situated the research community to tackle these challenges. Moreover, recent advances in campus network infrastructure, enabled largely by investments through the NSF CC* programs, provide opportunities not previously possible to create a highly federated distributed CISE research infrastructure.